\documentclass[10pt,a4paper,onecolumn]{article}
\usepackage{marginnote}
\usepackage{graphicx}
\usepackage[rgb]{xcolor}
\usepackage{authblk,etoolbox}
\usepackage{titlesec}
\usepackage{calc}
\usepackage{tikz}
\usepackage[pdfa]{hyperref}
\usepackage{hyperxmp}
\hypersetup{%
    unicode=true,
    pdfapart=3,
    pdfaconformance=B,
    pdftitle={UltraDark.jl: A Julia package for simulation of
cosmological scalar fields},
    pdfauthor={Nathan Musoke},
    pdfpublication={Journal of Open Source Software},
    pdfpublisher={Open Journals},
    pdfissn={2475-9066},
    pdfpubtype={journal},
    pdfvolumenum={},
    pdfissuenum={},
    pdfdoi={10.21105/joss.06035},
    pdfcopyright={Copyright (c) 2024, Nathan Musoke},
    pdflicenseurl={http://creativecommons.org/licenses/by/4.0/},
    colorlinks=true,
    linkcolor=[rgb]{0.0, 0.5, 1.0},
    citecolor=Blue,
    urlcolor=[rgb]{0.0, 0.5, 1.0},
    breaklinks=true
}
\immediate\pdfobj stream attr{/N 3} file{sRGB.icc}
\pdfcatalog{%
  /OutputIntents [
    <<
      /Type /OutputIntent
      /S /GTS_PDFA1
      /DestOutputProfile \the\pdflastobj\space 0 R
      /OutputConditionIdentifier (sRGB)
      /Info (sRGB)
    >>
  ]
}

\usepackage{caption}
\usepackage{orcidlink}
\usepackage{tcolorbox}
\usepackage{amssymb,amsmath}
\usepackage{ifxetex,ifluatex}
\usepackage{seqsplit}
\usepackage{xstring}

\usepackage{float}
\let\origfigure\figure
\let\endorigfigure\endfigure
\renewenvironment{figure}[1][2] {
    \expandafter\origfigure\expandafter[H]
} {
    \endorigfigure
}

\usepackage{fixltx2e} % provides \textsubscript

\newlength{\cslhangindent}
\newlength{\csllabelwidth}
\newenvironment{CSLReferences}[2] % #1 hanging-ident, #2 entry spacing
 {% don't indent paragraphs
  \setlength{\parindent}{0pt}
  \ifodd #1 \everypar{\setlength{\hangindent}{\cslhangindent}}\ignorespaces\fi
  \ifnum #2 > 0
  \setlength{\parskip}{#2\baselineskip}
  \fi
 }%
 {}
\usepackage{calc}

\usepackage[top=3.5cm, bottom=3cm, right=1.5cm, left=1.0cm,
            headheight=2.2cm, reversemp, includemp, marginparwidth=4.5cm]{geometry}

\titleformat{\section}
  {\normalfont\sffamily\Large\bfseries}
  {}{0pt}{}
\titleformat{\subsection}
  {\normalfont\sffamily\large\bfseries}
  {}{0pt}{}
\titleformat{\subsubsection}
  {\normalfont\sffamily\bfseries}
  {}{0pt}{}
\titleformat*{\paragraph}
  {\sffamily\normalsize}

\usepackage{fancyhdr}
\makeatletter
\let\ps@plain\ps@fancy
\fancyheadoffset[L]{4.5cm}
\fancyfootoffset[L]{4.5cm}

\definecolor{linky}{rgb}{0.0, 0.5, 1.0}

\newtcolorbox{repobox}
   {colback=red, colframe=red!75!black,
     boxrule=0.5pt, arc=2pt, left=6pt, right=6pt, top=3pt, bottom=3pt}

\newcommand{\ExternalLink}{%
   \tikz[x=1.2ex, y=1.2ex, baseline=-0.05ex]{%
       \begin{scope}[x=1ex, y=1ex]
           \clip (-0.1,-0.1)
               --++ (-0, 1.2)
               --++ (0.6, 0)
               --++ (0, -0.6)
               --++ (0.6, 0)
               --++ (0, -1);
           \path[draw,
               line width = 0.5,
               rounded corners=0.5]
               (0,0) rectangle (1,1);
       \end{scope}
       \path[draw, line width = 0.5] (0.5, 0.5)
           -- (1, 1);
       \path[draw, line width = 0.5] (0.6, 1)
           -- (1, 1) -- (1, 0.6);
       }
   }

\patchcmd{\@maketitle}{center}{flushleft}{}{}
\patchcmd{\@maketitle}{center}{flushleft}{}{}
\patchcmd{\@maketitle}{\LARGE}{\LARGE\sffamily}{}{}
\def\maketitle{{%
  
  \AB@maketitle}}
\renewcommand\AB@affilsepx{ \protect\Affilfont}
\renewcommand\AB@affilnote[1]{{\bfseries #1}\hspace{3pt}}
\renewcommand{\affil}[2][]%
   {\newaffiltrue\let\AB@blk@and\AB@pand
      \if\relax#1\relax\def\AB@note{\AB@thenote}\else\def\AB@note{#1}%
        \setcounter{Maxaffil}{0}\fi
        \begingroup
        \let\href=\href@Orig
        \let\protect\@unexpandable@protect
        \def\thanks{\protect\thanks}\def\footnote{\protect\footnote}%
        \@temptokena=\expandafter{\AB@authors}%
        {\def\\{\protect\\\protect\Affilfont}\xdef\AB@temp{#2}}%
         \xdef\AB@authors{\the\@temptokena\AB@las\AB@au@str
         \protect\\[\affilsep]\protect\Affilfont\AB@temp}%
         \gdef\AB@las{}\gdef\AB@au@str{}%
        {\def\\{, \ignorespaces}\xdef\AB@temp{#2}}%
        \@temptokena=\expandafter{\AB@affillist}%
        \xdef\AB@affillist{\the\@temptokena \AB@affilsep
          \AB@affilnote{\AB@note}\protect\Affilfont\AB@temp}%
      \endgroup
       \let\AB@affilsep\AB@affilsepx
}
\makeatother

\renewcommand\Affilfont{\sffamily\small\mdseries}
\ifnum 0\ifxetex 1\fi\ifluatex 1\fi=0 % if pdftex
  \usepackage[T1]{fontenc}
  \usepackage[utf8]{inputenc}

\else % if luatex or xelatex
  \ifxetex
    \usepackage{mathspec}
    \usepackage{fontspec}

  \else
    \usepackage{fontspec}
  \fi
  \defaultfontfeatures{Scale=MatchLowercase}
  \defaultfontfeatures[\sffamily]{Ligatures=TeX}
  \defaultfontfeatures[\rmfamily]{Ligatures=TeX,Scale=1}

\fi
\IfFileExists{upquote.sty}{\usepackage{upquote}}{}
\IfFileExists{microtype.sty}{%
\usepackage{microtype}
\UseMicrotypeSet[protrusion]{basicmath} % disable protrusion for tt fonts
}{}

\PassOptionsToPackage{usenames,dvipsnames}{color} % color is loaded by hyperref
\ifLuaTeX
\usepackage[bidi=basic]{babel}
\else
\usepackage[bidi=default]{babel}
\fi
\babelprovide[main,import]{american}
\def\languageshorthands#1{}

\usepackage{graphicx,grffile}
\makeatletter
\def\maxwidth{\ifdim\Gin@nat@width>\linewidth\linewidth\else\Gin@nat@width\fi}
\def\maxheight{\ifdim\Gin@nat@height>\textheight\textheight\else\Gin@nat@height\fi}
\makeatother
\setkeys{Gin}{width=\maxwidth,height=\maxheight,keepaspectratio}
\IfFileExists{parskip.sty}{%
\usepackage{parskip}
}{% else
\setlength{\parindent}{0pt}
\setlength{\parskip}{6pt plus 2pt minus 1pt}
}
\ifx\paragraph\undefined\else
\let\oldparagraph\paragraph
\renewcommand{\paragraph}[1]{\oldparagraph{#1}\mbox{}}
\fi
\ifx\subparagraph\undefined\else
\let\oldsubparagraph\subparagraph
\renewcommand{\subparagraph}[1]{\oldsubparagraph{#1}\mbox{}}
\fi
\ifLuaTeX
  \usepackage{selnolig}  % disable illegal ligatures
\fi

\title{UltraDark.jl: A Julia package for simulation of cosmological
scalar fields}

\author[%
]{Nathan Musoke%
  \,\orcidlink{0000-0001-9839-9256}\,%
}

\affil[]{Department of Physics and Astronomy, University of New
Hampshire, USA}
\date{\vspace{-2.5ex}}

\begin{document}
\maketitle

\marginpar{

  \begin{flushleft}
  %\hrule
  \sffamily\small

    {\bfseries Preprint:} Prepared for submission to Journal of Open
Source Software
  
  \vspace{2mm}
    {\bfseries Software}
  \begin{itemize}
    \setlength\itemsep{0em}
    \item \href{https://github.com/musoke/UltraDark.jl}{\color{linky}{Repository}} \ExternalLink
  \end{itemize}

  \vspace{2mm}

  \vspace{2mm}
  {\bfseries License}\\
  Authors of papers retain copyright and release the work under a Creative Commons Attribution 4.0 International License (\href{https://creativecommons.org/licenses/by/4.0/}{\color{linky}{CC BY 4.0}}).

  \end{flushleft}
}

\hypertarget{summary}{%
\section{Summary}\label{summary}}

\texttt{UltraDark.jl} is a Julia package for the simulation of
cosmological scalar fields. Scalar fields are proposed solutions to two
of the fundamental questions in cosmology: the nature of dark matter and
the universe's initial conditions. Modeling their dynamics requires
solving the Gross-Pitaevskii-Poisson equations, which is analytically
challenging. This makes simulations essential to understanding the
dynamics of cosmological scalar fields. \texttt{UltraDark.jl} is an
open, performant and user friendly option for solving these equations
numerically.

\begin{figure}
\centering
\includegraphics[width=1\textwidth,height=\textheight]{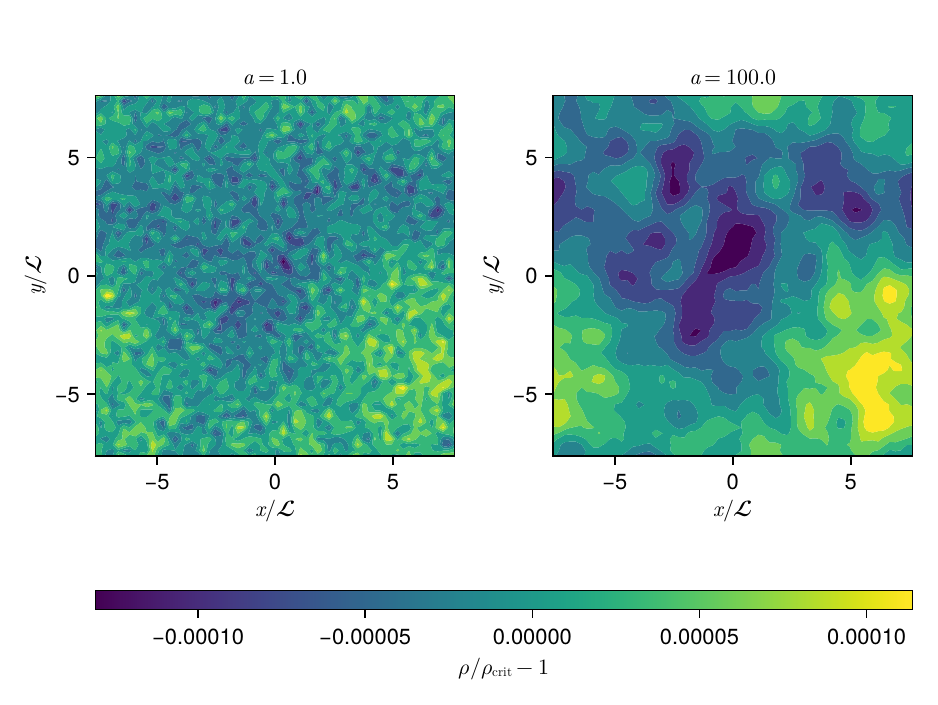}
\caption{Growth of density perturbations in a scalar field in an
expanding background, as the universe's scale factor growths from
\(a=1.0\) to \(a = 100.0\). The density is measured in units of the
universe's critical density \(\rho_{\mathrm{crit}}\). This scenario is
similar to that in (\protect\hyperlink{ref-Musoke:2019ima}{Musoke et
al., 2020}), in which the scalar field is an inflaton fragmenting after
the end of primordial inflation. \label{fig:growth}}
\end{figure}

\hypertarget{statement-of-need}{%
\section{Statement of need}\label{statement-of-need}}

Scalar fields are ubiquitous in physics, as solutions to partial
differential equations describing the spatial variation of physical
quantities. As dark matter candidates, scalar fields including
axion-like particles (ALPs) would explain the nature of the missing 85\%
of the universe's matter (\protect\hyperlink{ref-Adams:2022pbo}{Adams et
al., 2022}; \protect\hyperlink{ref-Planck:2015fie}{Ade \& others, 2016};
\protect\hyperlink{ref-Hu:2000ke}{Hu et al., 2000};
\protect\hyperlink{ref-Matos:1999et}{Matos et al., 2000}). As inflaton
candidates, scalar fields are proposed to cause a phase of accelerated
expansion that sets the stage for big bang nucleosynthesis
(\protect\hyperlink{ref-Albrecht:1982wi}{Albrecht \& Steinhardt, 1982};
\protect\hyperlink{ref-Amin:2011hj}{Amin et al., 2012};
\protect\hyperlink{ref-Guth:1980zm}{Guth, 1981};
\protect\hyperlink{ref-Linde:1981mu}{Linde, 1982}). In each case, a
scalar field \(\psi(t, \mathbf{x})\) represents the density of particles
as a function of space and time.

Subject to reasonable conditions, a cosmological scalar field \(\psi\)
whose particles have mass \(m\) obeys the Gross-Pitaevskii equation for
the scalar field, \begin{equation}
    \label{eq:gpp}
    i \hbar \frac{\partial \psi}{\partial t} = -\frac{\hbar^2}{2 m a(t)^2} \nabla^2 \psi + m \Phi \psi
    \;,
\end{equation} coupled to the Poisson equation for the gravitational
potential \(\Phi\) \begin{equation}
    \label{eq:poisson}
    \nabla^2 \Phi = \frac{4\pi G}{a(t)} m {|\psi|}^2
    \;,
\end{equation} where \(a(t)\) is the scale factor characterising the
expansion of the universe.

These equations are difficult to solve analytically -- even static
equilibrium solutions do not have a closed form -- and necessitate the
use of computer simulations. There are codes which solving
\autoref{eq:gpp} and \autoref{eq:poisson} with different methods and in
different domains, including \texttt{PyUltraLight}
(\protect\hyperlink{ref-Edwards:2018ccc}{Edwards et al., 2018}), a code
written in Chapel
(\protect\hyperlink{ref-padmanabhan2020simulating}{Padmanabhan et al.,
2020}), \texttt{AxioNyx}
(\protect\hyperlink{ref-Schwabe:2020eac}{Schwabe et al., 2020}), SCALAR
(\protect\hyperlink{ref-Mina:2019ekb}{Mina et al., 2020}), i-SPin 2
(\protect\hyperlink{ref-Jain:2023qty}{Jain et al., 2023};
\protect\hyperlink{ref-Jain:2022agt}{Jain \& Amin, 2023}); see Zhang et
al. (\protect\hyperlink{ref-Zhang:2018ghp}{2018}) for an overview.

\texttt{UltraDark.jl} solves \autoref{eq:gpp} and \autoref{eq:poisson}
with a pseudo-spectral symmetrized split-step method, in which each time
step consists of four sub-steps: \begin{gather}
    \label{eq:update_phase_1}
    \psi \to \exp\left( - i\frac{h}{2} \Phi \right) \psi
    \\
    \label{eq:update_density}
    \psi \to \mathcal{F}^{-1}\left\{ \exp\left(-i h \frac{k^2}{2} \right) \mathcal{F}\left\{ \psi \right\} \right\}
    \\
    \Phi = \mathcal{F}^{-1}\left\{ - \frac{4\pi}{a k^2} \mathcal{F}\left\{ |\psi|^2\right\}\right\}
    \\
    \label{eq:update_phase_2}
    \psi \to \exp\left( - i\frac{h}{2} \Phi \right) \psi
    ,
\end{gather} where \(\mathcal{F}\) is a Fourier transform, \(k\) are the
corresponding frequencies, and \(h\) is the time step.
\texttt{UltraDark.jl} has adaptive time steps which allow it to
accelerate simulations while preserving numerical convergence. This is
particularly useful in an expanding universe, where the time step is
roughly \(h \propto a^2\). Such time steps result in orders of magnitude
speedups when simulating collapse of an inflaton field in the early
universe (\protect\hyperlink{ref-Musoke:2019ima}{Musoke et al., 2020}).

\begin{figure}
\centering
\includegraphics[width=1\textwidth,height=\textheight]{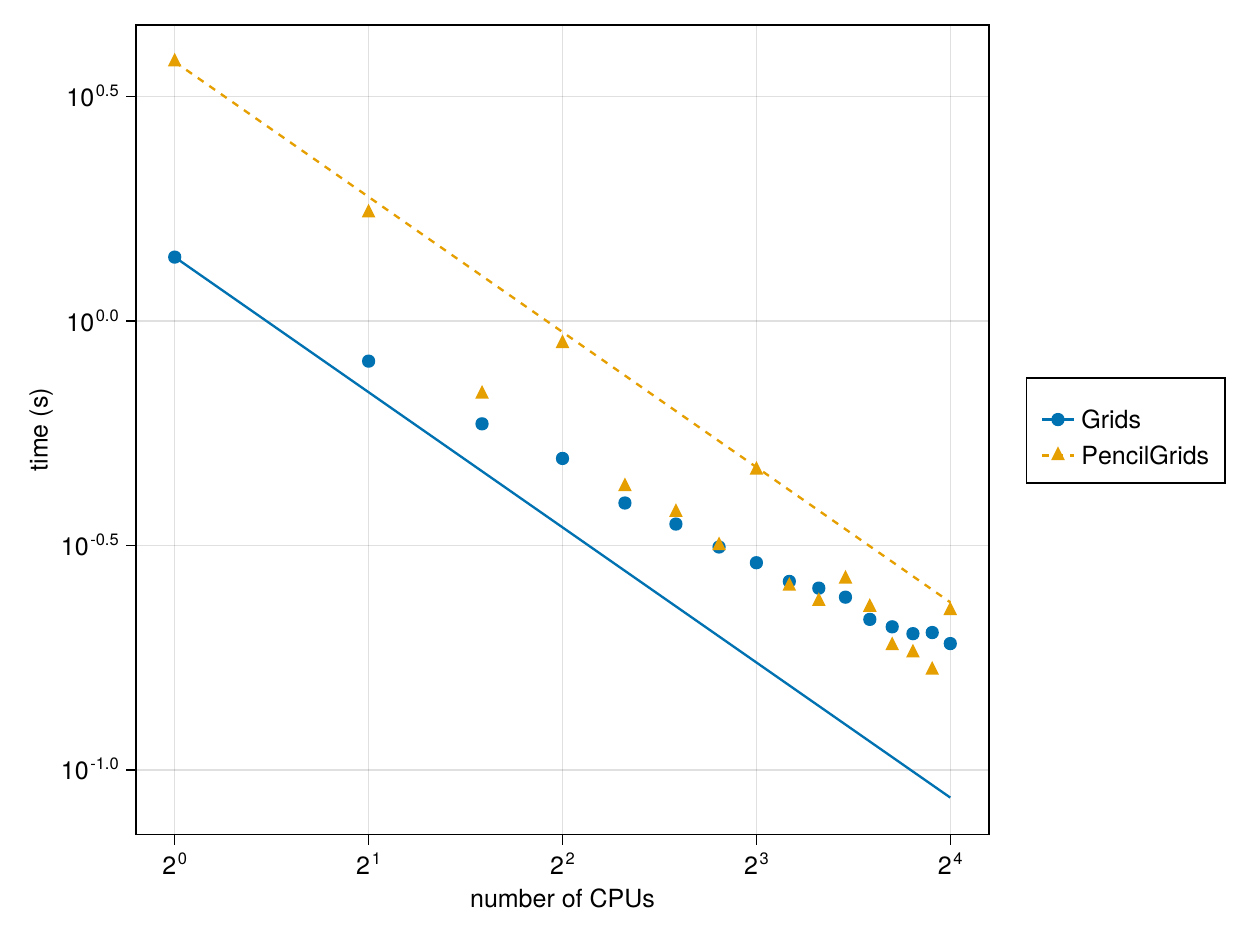}
\caption{Wall time for a single time step, as a function of number of
CPUs. The points represent measured times and the lines represent
theoretical \(1/\text{\#CPU}\) scalings. The circles and solid line are
for grids constructed from \texttt{Array}s and the triangles and dashed
lines are for MPI-distributed \texttt{PencilArray}s.\label{fig:cpus}}
\end{figure}

Julia (\protect\hyperlink{ref-Julia-2017}{Bezanson et al., 2017}) has
seen increasing use in scientific computing; see for example Eschle et
al. (\protect\hyperlink{ref-Eschle:2023ikn}{2023}) and Roesch et al.
(\protect\hyperlink{ref-JuliaBiologists}{2021}) for overviews of its use
in high energy physics and biology. The use of Julia is one of the
choices that separates \texttt{UltraDark.jl} from similar codes.
\texttt{UltraDark.jl} uses Julia's rich parallelism capabilities. The
\texttt{Threads.@threads} macro provides simple parallelisation of
\texttt{for} loops. Folds.jl enables simple parallelisation of reduction
operations (\protect\hyperlink{ref-folds}{Arakaki, 2020}). In a cluster
environment, \texttt{PencilArrays.jl} and \texttt{PencilFFTs.jl} enable
straightforward cross-node parallelism, a capability that is challenging
to reproduce in Python (\protect\hyperlink{ref-PencilArrays}{Polanco,
2021a}, \protect\hyperlink{ref-PencilFFTs}{2021b}). The scaling of this
two approaches to parallelism is demonstrated in \autoref{fig:cpus}.

The features described above have allowed collaborators and I to produce
results presented in 4 publications, each exploring an aspect of the
small scale structure of ultralight dark matter. We have used
\texttt{UltraDark.jl} to explore tidal disruption in dark matter halos
comprised of self-interacting ALPs
(\protect\hyperlink{ref-Glennon:2022huu}{Glennon et al., 2022}), the
effects of self-interactions on dynamical friction
(\protect\hyperlink{ref-Glennon:2023gfm}{Glennon et al., 2024}), perform
the first simulations of multi-species ALPs with intra- and
inter-species interactions
(\protect\hyperlink{ref-Glennon:2023jsp}{Glennon, Musoke, \&
Prescod-Weinstein, 2023b}) and discover a novel mechanism for vortex
stabilisation in scalar dark matter
(\protect\hyperlink{ref-Glennon:2023oqa}{Glennon, Mirasola, Musoke, et
al., 2023a}). More sample output can be found in Glennon, Mirasola,
Musoke, et al. (\protect\hyperlink{ref-anim:vortex}{2023b}), Glennon,
Musoke, \& Prescod-Weinstein
(\protect\hyperlink{ref-anim:multi}{2023a}), and Glennon, Musoke,
Nadler, et al. (\protect\hyperlink{ref-anim:friction}{2023}).

\hypertarget{acknowledgements}{%
\section{Acknowledgements}\label{acknowledgements}}

I thank Arka Banerjee, Ben Chandran, Richard Easther, Mateja Gosenca,
Shaun Hotchkiss, Emily Kendall, Anthony Mirasola, Ethan Nadler, Mark
Neyrinck, Chanda Prescod-Weinstein, Yourong Frank Wang, Risa Wechsler,
Luna Zagorac for discussion and comments on various stages of the
development of \texttt{UltraDark.jl}. Particular thanks to Noah Glennon
as the first user of \texttt{UltraDark.jl}.

Computations were performed on Marvin, a Cray CS500 supercomputer at UNH
supported by the NSF MRI program under grant AGS-1919310. The author
wishes to acknowledge the use of New Zealand eScience Infrastructure
(NeSI) high performance computing facilities, consulting support and/or
training services as part of this research. New Zealand's national
facilities are provided by NeSI and funded jointly by NeSI's
collaborator institutions and through the Ministry of Business,
Innovation \& Employment's Research Infrastructure programme. This work
was performed in part at Aspen Center for Physics, which is supported by
National Science Foundation under Grant No.~PHY-1607611. This work was
partially supported by a grant from the Sloan Foundation.

\hypertarget{references}{%
\section*{References}\label{references}}
\addcontentsline{toc}{section}{References}

\hypertarget{refs}{}
\begin{CSLReferences}{1}{0}
\leavevmode\vadjust pre{\hypertarget{ref-Adams:2022pbo}{}}%
Adams, C. B., Aggarwal, N., Agrawal, A., Balafendiev, R., Bartram, C.,
Baryakhtar, M., Bekker, H., Belov, P., Berggren, K. K., Berlin, A.,
Boutan, C., Bowring, D., Budker, D., Caldwell, A., Carenza, P., Carosi,
G., Cervantes, R., Chakrabarty, S. S., Chaudhuri, S., \ldots{} Zhou, T.
(2022, March). Axion dark matter. \emph{{Snowmass 2021}}.
\url{https://doi.org/10.48550/ARXIV.2203.14923}

\leavevmode\vadjust pre{\hypertarget{ref-Planck:2015fie}{}}%
Ade, P. A. R., \& others. (2016). {Planck 2015 results. XIII.
Cosmological parameters}. \emph{Astron. Astrophys.}, \emph{594}, A13.
\url{https://doi.org/10.1051/0004-6361/201525830}

\leavevmode\vadjust pre{\hypertarget{ref-Albrecht:1982wi}{}}%
Albrecht, A., \& Steinhardt, P. J. (1982). {Cosmology for Grand Unified
Theories with Radiatively Induced Symmetry Breaking}. \emph{Phys. Rev.
Lett.}, \emph{48}, 1220--1223.
\url{https://doi.org/10.1103/PhysRevLett.48.1220}

\leavevmode\vadjust pre{\hypertarget{ref-Amin:2011hj}{}}%
Amin, M. A., Easther, R., Finkel, H., Flauger, R., \& Hertzberg, M. P.
(2012). {Oscillons After Inflation}. \emph{Phys. Rev. Lett.},
\emph{108}, 241302. \url{https://doi.org/10.1103/PhysRevLett.108.241302}

\leavevmode\vadjust pre{\hypertarget{ref-folds}{}}%
Arakaki, T. (2020). Folds: Sequential, threaded, and distributed fold
interface for julia. In \emph{GitHub repository}. GitHub.
\url{https://github.com/JuliaFolds/Folds.jl}

\leavevmode\vadjust pre{\hypertarget{ref-Julia-2017}{}}%
Bezanson, J., Edelman, A., Karpinski, S., \& Shah, V. B. (2017). Julia:
A fresh approach to numerical computing. \emph{SIAM {R}eview},
\emph{59}(1), 65--98. \url{https://doi.org/10.1137/141000671}

\leavevmode\vadjust pre{\hypertarget{ref-Edwards:2018ccc}{}}%
Edwards, F., Kendall, E., Hotchkiss, S., \& Easther, R. (2018).
PyUltraLight: A pseudo-spectral solver for ultralight dark matter
dynamics. \emph{JCAP}, \emph{10}, 027.
\url{https://doi.org/10.1088/1475-7516/2018/10/027}

\leavevmode\vadjust pre{\hypertarget{ref-Eschle:2023ikn}{}}%
Eschle, J., Gal, T., Giordano, M., Gras, P., Hegner, B., Heinrich, L.,
Acosta, U. H., Kluth, S., Ling, J., Mato, P., Mikhasenko, M., Briceño,
A. M., Pivarski, J., Samaras-Tsakiris, K., Schulz, O., Stewart, G. A.,
Strube, J., \& Vassilev, V. (2023). Potential of the julia programming
language for high energy physics computing. \emph{Computing. Comput
Softw Big Sci 7, 10 (2023)}, \emph{7}(1).
\url{https://doi.org/10.1007/s41781-023-00104-x}

\leavevmode\vadjust pre{\hypertarget{ref-Glennon:2023oqa}{}}%
Glennon, N., Mirasola, A. E., Musoke, N., Neyrinck, M. C., \&
Prescod-Weinstein, C. (2023a). Scalar dark matter vortex stabilization
with black holes. \emph{Journal of Cosmology and Astroparticle Physics},
\emph{2023}(07), 004.
\url{https://doi.org/10.1088/1475-7516/2023/07/004}

\leavevmode\vadjust pre{\hypertarget{ref-anim:vortex}{}}%
Glennon, N., Mirasola, A. E., Musoke, N., Neyrinck, M. C., \&
Prescod-Weinstein, C. (2023b). \emph{{Supplementary animations for
"Scalar dark matter vortex stabilization with black holes"}}. Zenodo.
\url{https://doi.org/10.5281/zenodo.7675830}

\leavevmode\vadjust pre{\hypertarget{ref-anim:friction}{}}%
Glennon, N., Musoke, N., Nadler, E. O., Prescod-Weinstein, C., \&
Wechsler, R. H. (2023). \emph{{Supplementary animations for "Dynamical
friction in self-interacting ultralight dark matter"}}. Zenodo.
\url{https://doi.org/10.5281/zenodo.7927474}

\leavevmode\vadjust pre{\hypertarget{ref-Glennon:2023gfm}{}}%
Glennon, N., Musoke, N., Nadler, E. O., Prescod-Weinstein, C., \&
Wechsler, R. H. (2024). {Dynamical friction in self-interacting
ultralight dark matter}. \emph{Phys. Rev. D}, \emph{109}(6), 063501.
\url{https://doi.org/10.1103/PhysRevD.109.063501}

\leavevmode\vadjust pre{\hypertarget{ref-anim:multi}{}}%
Glennon, N., Musoke, N., \& Prescod-Weinstein, C. (2023a).
\emph{{Supplementary animations for "Simulations of multi-field
ultralight axion-like dark matter"}}. Zenodo.
\url{https://doi.org/10.5281/zenodo.7675775}

\leavevmode\vadjust pre{\hypertarget{ref-Glennon:2023jsp}{}}%
Glennon, N., Musoke, N., \& Prescod-Weinstein, C. (2023b). Simulations
of multi-field ultralight axion-like dark matter. \emph{Physical Review
D}, \emph{107}(6), 063520.
\url{https://doi.org/10.1103/physrevd.107.063520}

\leavevmode\vadjust pre{\hypertarget{ref-Glennon:2022huu}{}}%
Glennon, N., Nadler, E. O., Musoke, N., Banerjee, A., Prescod-Weinstein,
C., \& Wechsler, R. H. (2022). Tidal disruption of solitons in
self-interacting ultralight axion dark matter. \emph{Physical Review D},
\emph{105}(12), 123540.
\url{https://doi.org/10.1103/physrevd.105.123540}

\leavevmode\vadjust pre{\hypertarget{ref-Guth:1980zm}{}}%
Guth, A. H. (1981). {The Inflationary Universe: A Possible Solution to
the Horizon and Flatness Problems}. \emph{Phys. Rev. D}, \emph{23},
347--356. \url{https://doi.org/10.1103/PhysRevD.23.347}

\leavevmode\vadjust pre{\hypertarget{ref-Hu:2000ke}{}}%
Hu, W., Barkana, R., \& Gruzinov, A. (2000). Cold and fuzzy dark matter.
\emph{Phys. Rev. Lett.}, \emph{85}, 1158--1161.
\url{https://doi.org/10.1103/PhysRevLett.85.1158}

\leavevmode\vadjust pre{\hypertarget{ref-Jain:2022agt}{}}%
Jain, M., \& Amin, M. A. (2023). {i-SPin}: An integrator for
multicomponent schrödinger-poisson systems with self-interactions.
\emph{JCAP}, \emph{04}, 053.
\url{https://doi.org/10.1088/1475-7516/2023/04/053}

\leavevmode\vadjust pre{\hypertarget{ref-Jain:2023qty}{}}%
Jain, M., Amin, M. A., \& Pu, H. (2023). {i-SPin 2}: An integrator for
general spin-s gross-pitaevskii systems. \emph{Phys. Rev. E 108, 055305,
15 November 2023}, \emph{108}(5), 055305.
\url{https://doi.org/10.1103/physreve.108.055305}

\leavevmode\vadjust pre{\hypertarget{ref-Linde:1981mu}{}}%
Linde, A. D. (1982). {A New Inflationary Universe Scenario: A Possible
Solution of the Horizon, Flatness, Homogeneity, Isotropy and Primordial
Monopole Problems}. \emph{Phys. Lett. B}, \emph{108}, 389--393.
\url{https://doi.org/10.1016/0370-2693(82)91219-9}

\leavevmode\vadjust pre{\hypertarget{ref-Matos:1999et}{}}%
Matos, T., Guzman, F. S., \& Urena-Lopez, L. A. (2000). Scalar field as
dark matter in the universe. \emph{Class. Quant. Grav.}, \emph{17},
1707--1712. \url{https://doi.org/10.1088/0264-9381/17/7/309}

\leavevmode\vadjust pre{\hypertarget{ref-Mina:2019ekb}{}}%
Mina, M., Mota, D. F., \& Winther, H. A. (2020). SCALAR: An AMR code to
simulate axion-like dark matter models. \emph{Astron. Astrophys.},
\emph{641}, A107. \url{https://doi.org/10.1051/0004-6361/201936272}

\leavevmode\vadjust pre{\hypertarget{ref-Musoke:2019ima}{}}%
Musoke, N., Hotchkiss, S., \& Easther, R. (2020). Lighting the dark:
Evolution of the postinflationary universe. \emph{Phys. Rev. Lett.},
\emph{124}(6), 061301.
\url{https://doi.org/10.1103/PhysRevLett.124.061301}

\leavevmode\vadjust pre{\hypertarget{ref-padmanabhan2020simulating}{}}%
Padmanabhan, N., Ronaghan, E., Zagorac, J. L., \& Easther, R. (2020).
Simulating ultralight dark matter in chapel. \emph{2020 IEEE
International Parallel and Distributed Processing Symposium Workshops
(IPDPSW)}, 678--678.
\url{https://doi.org/10.1109/ipdpsw50202.2020.00120}

\leavevmode\vadjust pre{\hypertarget{ref-PencilArrays}{}}%
Polanco, J. I. (2021a). \emph{{PencilArrays.jl: Distributed Julia arrays
using the MPI protocol}} (Version 0.9.9).
\url{https://doi.org/10.5281/zenodo.5148035}

\leavevmode\vadjust pre{\hypertarget{ref-PencilFFTs}{}}%
Polanco, J. I. (2021b). \emph{{PencilFFTs.jl: FFTs of MPI-distributed
Julia arrays}} (Version 0.12.4).
\url{https://doi.org/10.5281/zenodo.3618781}

\leavevmode\vadjust pre{\hypertarget{ref-JuliaBiologists}{}}%
Roesch, E., Greener, J. G., MacLean, A. L., Nassar, H., Rackauckas, C.,
Holy, T. E., \& Stumpf, M. P. H. (2021). \emph{Julia for biologists}.
\url{https://doi.org/10.48550/ARXIV.2109.09973}

\leavevmode\vadjust pre{\hypertarget{ref-Schwabe:2020eac}{}}%
Schwabe, B., Gosenca, M., Behrens, C., Niemeyer, J. C., \& Easther, R.
(2020). Simulating mixed fuzzy and cold dark matter. \emph{Phys. Rev.
D}, \emph{102}(8), 083518.
\url{https://doi.org/10.1103/PhysRevD.102.083518}

\leavevmode\vadjust pre{\hypertarget{ref-Zhang:2018ghp}{}}%
Zhang, J., Liu, H., \& Chu, M.-C. (2018). Cosmological simulation for
fuzzy dark matter model. \emph{Frontiers in Astronomy and Space
Sciences}, \emph{5}. \url{https://doi.org/10.3389/fspas.2018.00048}

\end{CSLReferences}

\end{document}